\documentclass{IEEEtran}
\usepackage{cite}
\usepackage{amsmath,amssymb,amsfonts}
\usepackage{algorithmic}
\usepackage{graphicx}
\usepackage{textcomp}
\usepackage{epstopdf}
\usepackage{bm}
\usepackage{float}
\usepackage{threeparttable}
\usepackage{caption}
\usepackage{array}
\usepackage{setspace}
\usepackage{gensymb}
\usepackage{stfloats}

\def\BibTeX{{\rm B\kern-.05em{\sc i\kern-.025em b}\kern-.08em
    T\kern-.1667em\lower.7ex\hbox{E}\kern-.125emX}}
\begin{document}

\title{Data-driven Discovery of Partial Differential Equations for Multiple-Physics Electromagnetic Problem}
\author{Bing Xiong, Haiyang Fu, Feng Xu, Yaqiu Jin}

%Key Laboratory for Information Science of %Electromagnetic Waves (Ministry of Education), School %of Information Science and Engineering, Fudan %University, Shanghai}

\maketitle

\begin{abstract}
Deriving governing equations in Electromagnetic (EM) environment based on first principles can be quite tough when there are some unknown sources of noise and other uncertainties in the system. For nonlinear multiple-physics electromagnetic systems, deep learning to solve these problems can achieve high efficiency and accuracy. In this paper, we propose a deep learning neutral network in combination with sparse regression to solve the hidden governing equations in multiple-physics EM problem. Pareto analysis is also adopted to preserve inversion as precise and simple as possible. This proposed network architecture can discover a set of governing partial differential equations (PDEs) based on few temporal-spatial samples. The data-driven discovery method for partial differential equations (PDEs) in electromagnetic field may also contribute to solve more sophisticated problem which may not be solved by first principles.
\end{abstract}

\begin{IEEEkeywords}
deep learning, convolutional neural network, partial differential equations, multiple electromagnetic physics;
\end{IEEEkeywords}

\section{Introduction}
\label{sec:introduction}
With comprehensive application of electromagnetic wave spectrum in radar, communication and navigation and increased computing power and storage, the amount of electromagnetic data comes to a huge number. Less well understood is how to distill the underlying electromagnetic physical laws from electromagnetic data in complex systems. The big electromagnetic data provides new opportunities for the data-driven discovery of new physics laws or making complex system modelling computationally feasible. Traditional derivation of governing equations relies on fundamental laws and theoretical  solution can be obtained with analytic and computational methods. However, the realistic scene in application is complex to tackle, which involves multiple-physics, multiple-scale and nonlinear. For a multiple-physics system with complex interaction mechanisms, there is no exact quantitative analytic solution and computational cost is high to solve a set of partial differential equations (PDEs). Therefore, the key question is how to use sparse data given to discover the principle of a complex system to make predictions of physical quantities outside the range of the measurement data. If the model can be learned from spare data and then perform prediction, it is significant for complex electromagnetic multiple-physics system and beyond.

The data-driver discovery methods have been developing quickly during the past decades. It can be mainly divided into three categories, including symbolic regression, sparse regression and deep learning. One of earlier research on data-driven discovery for free-form natural laws is proposed by Bongard and Lipson (2007) and Schmidt and Lipson (2009) \cite{b1}. The main idea is to calculate numerical differentiation of experimental data first and use symbolic regression based on evolutionary algorithm to compare with analytical derivative solutions. Later, sparse regression method was raised by Brunton et al. (2016) \cite{b3}, Schaeffer (2017)\cite{b4}, and Rudy et al. (2017)\cite{b5}. A candidate function dictionary needs to be established which consists of simple and derivatives terms and then use sparse regression to select the proper terms as part of the equations. Recently, Raissi and Karniadakis (2017) used full connection deep neural network to discover underlying physics of nonlinear PDEs \cite{b6,b7} with less data. The key idea is to set up a universal neural network to approximate the solution of the PDE by minimizing the loss function and the derivatives can then be calculated by automation differentiation based on neural network\cite{b7}. However, the explicit form of the PDEs is assumed to be known. Recently, Long et al. (2018) \cite{b7} utilized the connection between differentiation and convolution to discover nonlinear equations with minor knowledge on the equation form. The wavelet frame filters in convolutional neural network (CNN) with a training kernel is adopted to approximate spatial differentiations. Long et al. (2018) \cite{b8} upgraded their network by imposing appropriate constraints on filters and using a newly designed symbolic neural network to express the analytical form of function clearly.

In sum, the data-driven discovery method of the underlying physics of PDEs are in progress. However, they still have some issues, such as symbolic regression with high computation cost and difficulty for dealing with large scale problem. Sparse regression needs to set numerical differentiations beforehand which costs large storage and may generate unrelated terms. The PDE-net for discovering unknown equations by Long et al. (2018) \cite{b6}\cite{b7} requires no knowledge on the differential operators and associated discrete approximations, which has been applied into 2-dimensional linear variable-coefficient convection-diffusion equation.

However, it is still challenging for the data-driven method in order to solve multiple-physics electromagnetic physics problem. There are two main characteristics for data-driven methods for multiple physics electromagnetic application. First, the electromagnetic scatter field is usually varying fast with time. The characteristic of the EM field needs to be extracted in the time series of data. Secondly,  multiple-physics electromagnetic problems involve a set of coupled different partial differential equations, which may inherit time-varying inhomogeneous coefficients. It requires one unified algorithm to retrieve multiple coefficients of the PDE equation set simultaneously. Therefore, the data-driven discovery method for complex electromagnetic system will be specially designed as a framework.

This paper aims to design a data-driven network architecture to discover a set of nonlinear PDE equations for multiple-physics EM problem. A unified neutral network with CNN is proposed in combination with spare regression and pareto analysis. This proposed network is applied for a electromagnetic wave and plasma interaction system, which can discover the coefficient of the PDE set with relative good accuracy. Also, we investigate the possibility of inferring inhomogeneous coefficients with little prior knowledge. The data-driven methods of the PDEs set will pay the way for deriving equations for complex partially known and unknown systems, including nonlinear, multiple physics, Maxwell’s equations equations and beyond.
% The following content will be organized as introduction, methodology and architecture, simulation and results, conclusion and discussion.

\section{Methodology}
For simplicity, we assume there are two physical variable $ u(x,y,t), v(x,y,t) $ in a 2-dimensional space $(x,y)$ varying with time $t$. The universal expression of their governing equations can be expressed as following
%具体的定义要写清楚，u要说明一下

\begin{spacing}{2.0}
\begin{equation}
\begin{cases}
\dfrac{\partial u}{\partial t} = f(x,y,u,v,\dfrac{\partial u}{\partial x},\dfrac{\partial u}{\partial y},\dfrac{\partial v}{\partial x},\dfrac{\partial v}{\partial y},\dfrac{\partial^{2} u}{\partial x \partial y},\dfrac{\partial^{2} v}{\partial x \partial y}...)\\

\dfrac{\partial v}{\partial t} = g(x,y,u,v,\dfrac{\partial v}{\partial x},\dfrac{\partial v}{\partial y},\dfrac{\partial u}{\partial x},\dfrac{\partial u}{\partial y},\dfrac{\partial^{2} v}{\partial x \partial y},\dfrac{\partial^{2} u}{\partial x \partial y}...)\label{eq}

\end{cases}
\end{equation}
\end{spacing}
where $f$ and $g$ are  nonlinear function of all possible terms including $u$ and $v$, their first and second order derivatives and other parameters, respectively.
%As both equations have equal state in this equation set, we take function $ u $ as example to introduce math process, the process is totally same for function $ v $.
\subsection{Time derivative}
We define $ u(t_{i},\cdot) $ or $ v(t_{i},\cdot) $ as all spatial value of function $ u $ or $v$ at $ t = t_{i} $. For time derivative, according to the forward euler method, we obtain the following formula

\begin{equation}
\begin{split}
\tilde{u}(t_{i+1},\cdot)=u(t_{i},\cdot)+\Delta t\cdot\dfrac{\partial u}{\partial t}\\
\tilde{v}(t_{i+1},\cdot)=v(t_{i},\cdot)+\Delta t\cdot\dfrac{\partial v}{\partial t}
\end{split}
\label{eq}
\end{equation}

where $ \tilde{u}(t_{i+1},\cdot) $ or $ \tilde{v}(t_{i+1},\cdot) $ is the approximation value at $ t_{i+1} $.
Then, we obtain expression

\begin{equation}
\begin{split}
\tilde{u}(t_{i+1},\cdot)=u(t_{i},\cdot)+\Delta t\cdot f\\
\tilde{v}(t_{i+1},\cdot)=v(t_{i},\cdot)+\Delta t\cdot g
\end{split}
\label{eq}
\end{equation}

\subsection{Spatial derivative}
For spatial derivative calculation, the connection between convolutions and differentiations was studies by Cai et al. (2012) \cite{b10} and by Dong et al. (2017) \cite{b8}. Here, we demonstrate one simple example of their work \cite{b8}\cite{b10} to show how to express differentiation by convolution. Consider that the 2-dimensional Haar wavelet frame filter banks containing one low-pass filter $ h_{00} $ and three high-pass filters $ h_{10} $, $ h_{01} $ and $ h_{11} $ have forms :

\begin{equation}
\begin{split}
h_{00}=\dfrac{1}{4}\begin{pmatrix}1 & 1\\1 & 1\end{pmatrix},h_{10}=\dfrac{1}{4}\begin{pmatrix}1 & -1\\1 & -1\end{pmatrix},\\
h_{01}=\dfrac{1}{4}\begin{pmatrix}1 & 1\\-1 & -1\end{pmatrix},h_{11}=\dfrac{1}{4}\begin{pmatrix}1 & -1\\-1 & 1\end{pmatrix}.
\end{split}
\label{eq}
\end{equation}

Then, the circular convolution (labeled as $\otimes$ ) between filters and function $u$ for instance is expressed as:
\begin{equation}
\begin{split}
h_{00}\otimes u\approx {u}, \\
h_{10}\otimes u\approx \dfrac{1}{2}\delta_{x}\dfrac{\partial u}{\partial x}, \\h_{01}\otimes u\approx \dfrac{1}{2}\delta_{y}\dfrac{\partial u}{\partial y}, \\h_{11}\otimes u\approx \dfrac{1}{4}\delta_{x}\delta_{y}\dfrac{\partial^{2} u}{\partial x \partial y}.
\end{split}
\label{eq}
\end{equation}

where $ \delta_{x} $ and $ \delta_{y}$ are the spatial size of $x$ and $y$ direction of function $ u $ or $ v $, respectively. \\
Similarly for $v$,
\begin{equation}
\begin{split}
h_{00}\otimes v\approx {v}, \\
h_{10}\otimes v\approx \dfrac{1}{2}\delta_{x}\dfrac{\partial v}{\partial x}, \\h_{01}\otimes v\approx \dfrac{1}{2}\delta_{y}\dfrac{\partial v}{\partial y}, \\h_{11}\otimes v\approx \dfrac{1}{4}\delta_{x}\delta_{y}\dfrac{\partial^{2} v}{\partial x \partial y}.
\end{split}
\label{eq}
\end{equation}

By multiplying constant coefficients, convolution can represent differentiation in neural network effectively. The calculation of the second order differentiation is also mentioned in \cite{b7}, it is ignored since our current work does not involve it  as will be discussed in future.

\subsection{$ \Delta t $ block}

Therefore, we obtain a partial neural network structure from time derivative calculation based on forward Euler's method. For simplicity, the partial neural network architecture will be shown only for variable $u$ and target function $f$, similarly for variable $v$ and target function $g$ .

Figure 1 illustrates a $ \Delta t $ block \cite{b8} as a layer of neural network to advance variables based on equations in time. In principle, the structure of the $\Delta t $ block is an interpretation of Equation (3). The two left terms in Fig.1 represent $\Delta t \cdot f $, that is a candidate terms database in the form of neural network. Thus, the input of this network is clear, that's all possible functions with $ u $ and $ v $ in this example. Here, it is noted that for variable $v$ and its derivatives are also included for advancing $u$ in the candidate database because we assume that there is coupling between $u$ and $v$. The prior knowledge will help reduce the number of possible candidate functions associated with PDEs. The output of $ \delta t $ block is the predicted value at the next time stamp $ \tilde{u}(t_{i+1},\cdot) $. 

In general, the whole network consists of several $\Delta t$ blocks by continuously connecting each block one by one. The number of $ \Delta t $ blocks is determined by different convergence criteria.  

%111111111111111111111111111111111
%111111111111111111111111111111111
%111111111111111111111111111111111
\begin{figure*}[htbp]
\centering
\includegraphics[width=0.8\textwidth]{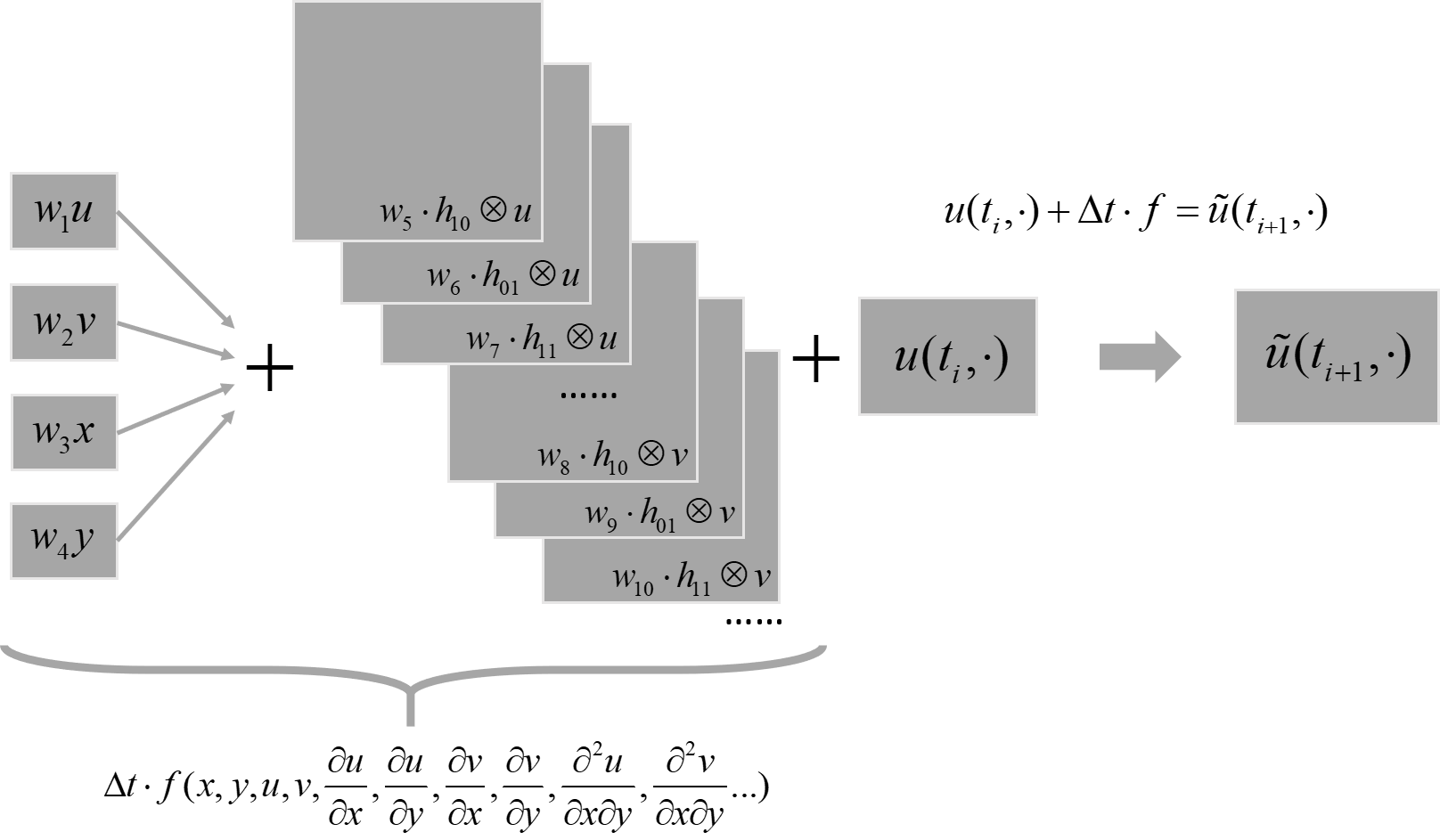}
\caption{The architecture of a $ \Delta t $ block \cite{b6} to advance function in time. The two left parts represent the target function $f$ in Equation (3), that includes a candidate term database in the form of neural network as in Equation (5-6). The input of this network is all possibly related functions which are $u$ and $v$ in this example. It is noted that function $v$ and its derivatives are also included in the candidate database because we assume that there exsits coupling between $u $ and $v$ in PDEs. }
\end{figure*}

%这段应该放在总体网络介绍之后，loss介绍之前
In Fig. 1, the coefficients $w_{1},w_{2},w_{3}...$ are neural network weights for each candidate term, respectively. The network weights will be shared in each layer and become the output vector of this model. The non-zero weight represents the term of the target equation, where $n$ means the number of candidate terms. We define it $\textbf{W}$ as
\begin{equation}
\textbf{W}= [w_{1}, w_{2}, w_{3}, w_{4}, w_{5},...,w_{n}],
\label{eq}
\end{equation}

\subsection{Sparse regression}
Sparse regression is a common method in data-discovery area. When there are many terms to discover with spare data, it means that the system may face with over-fitting problems. Regularization method need to be adopted in training process and $ L_{1} $ norm regularization is commonly used as $L_{1}$ is easier to calculate than $ L_{0} $ norm regularization.

In our approach, when the training of network is finished, the output $ \textbf{W} $ will be limited by regularization method in the final layer of network.

The output of $ \delta t $ block is the predicted value at the next time step $ \tilde{u}(t_{i+1},\cdot) $. Therefore, the loss function of a single block is defined as
\begin{equation}
l_{i} = \parallel \tilde{u}(t_{i+1},\cdot)-u(t_{i+1},\cdot) \parallel _{2}
\label{eq}
\end{equation}

Consider sparse regression to optimize the output $ \textbf{W} $ at the end of the complete network, we define that loss function of the entire network is sum of each block's loss according to $L_{1}$ norm of $\textbf{W}$. The accumulated loss function can make prediction reliable within certain time step numbers $N$. The expression for final loss $L$ is 
\begin{equation}
L = \sum_{i=1}^N \parallel \tilde{u}(t_{i},\cdot)-u(t_{i},\cdot) \parallel _{2}+\lambda\parallel \textbf{W} \parallel_{1}
\label{eq}
\end{equation}
where $\lambda$ is the regularization parameter to determine the weight of regularization term.

\subsection{Pareto analysis}
Pareto analysis is a technique that can be useful in situations where many possible action processes are competing for attention. In essence, the problem-solver evaluates the benefits of each action and selects some of the most effective actions that bring the total benefits reasonably close to the maximum possible benefits. While it is common to refer to pareto as the "80/20" rule, in all cases, 20\% of the reasons for 80\% of the problems are assumed to be a convenient rule of thumb, not and should not be considered an immutable law of nature. This technique helps identify the top priority part of the problem that needs to be addressed.

The application of Pareto analysis adopted in \cite{b6} is the final guarantee of the discovery result of the system. By cutting the term with the minimum parameter out and training again, we compare the later loss with the previous one. If loss reduces, then we repeat the process until loss rises.

Figure 2 illustrates the architecture of the whole neutral network to deal with a given PDE equation set. This process includes 
\begin{enumerate}
  \item Find out the number of possible equations to discover, e. g., $f$ or $g$ .
  \item Determine time derivatives, e. g.,$\tilde{u}(t_{i+1},\cdot)$ or $\tilde{v}(t_{i+1},\cdot)$.
  \item Decide the number of networks to activate.
  \item Utilize spare regression to train network weights $\textbf{W}$.
  \item Cut out terms based on Pareto analysis.
  \item Iterate and train again if not converged.
\end{enumerate}

%22222222222222222222222222222222
%22222222222222222222222222222222
%22222222222222222222222222222222
\begin{figure*}[htbp]
\centering
\includegraphics[width=0.7\textwidth]{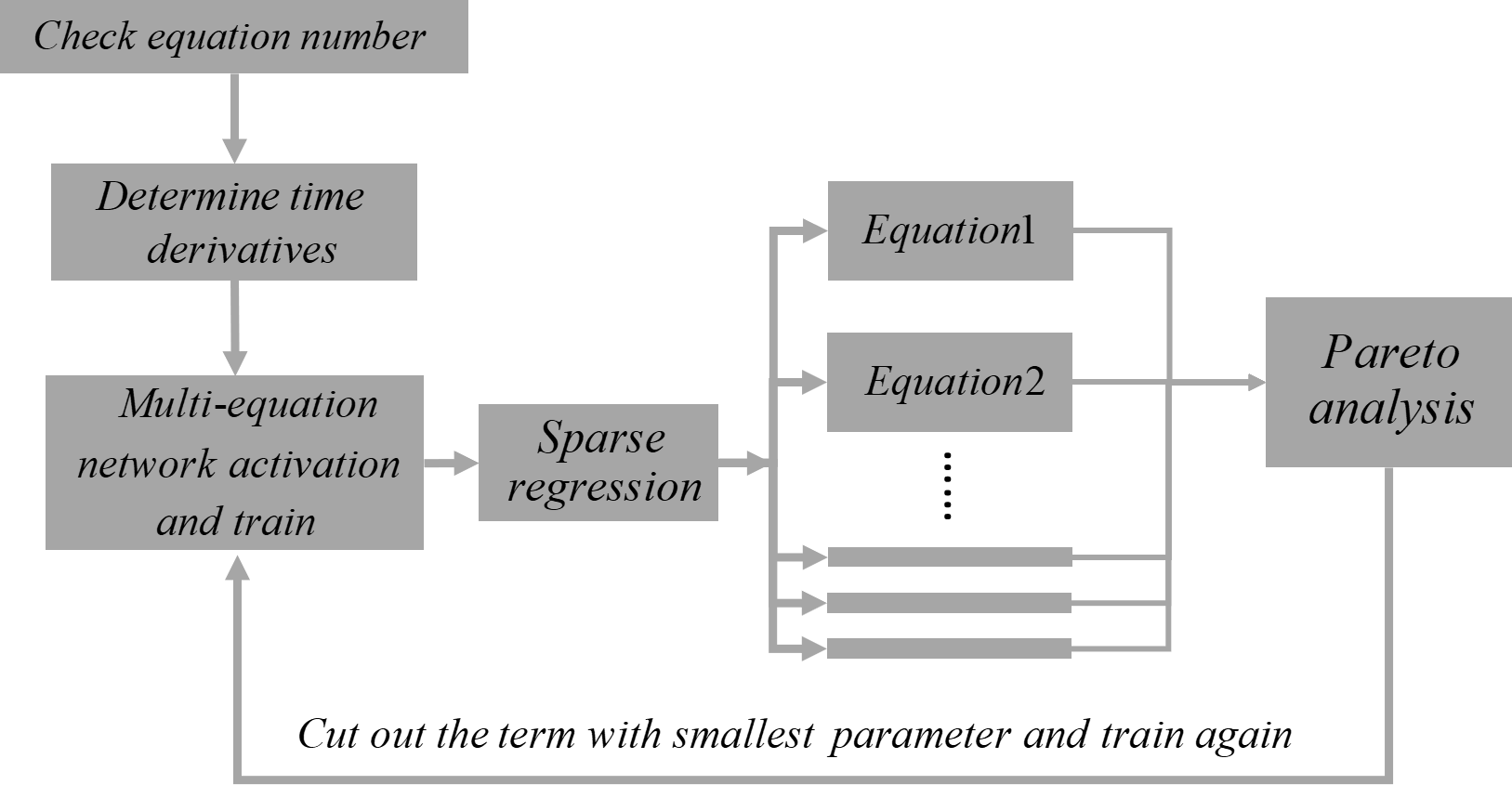}
\caption{The architecture of discovering the PDE equation set. The procedure steps include 1) Check the number of PDEs equations. 2) Determine the time derivative of active terms in the PDE. 3) Activate multiple equation network corresponding to each equations. 4) Train network with sparse regression. 5) Pareto analysis helps to cut out unrelated terms with minimum parameters.}
\end{figure*}

\section{Simulation}
\subsection{Forward model}
Here, we consider the interaction between electromagnetic waves and magnetized plasmas with collisions. For such multiple physic complex system, the Maxwell's partial differential equations and constitutive relations are given as

\begin{equation}
\begin{split}
\nabla\times\textbf{H}=\varepsilon_{0}\dfrac{\partial \textbf{E}}{\partial t}+\textbf{J}\\
\nabla\times\textbf{E}=-\mu_{0}\dfrac{\partial \textbf{H}}{\partial t}\\
\dfrac{\partial \textbf{J}}{\partial t}+\nu_{c}\textbf{J}=\varepsilon_{0}\omega_{p}^{2}\textbf{E}+\omega_{ce}\times\textbf{J}
\end{split}
\end{equation}

where ${\mathbf{H}}$ , ${\mathbf{E}}$ and ${\mathbf{J}}$ are magnetic intensity vector, electric field intensity vector, and polarized current density vector respectively. ${\varepsilon _0}$, ${\mu _0}$ are the vacuum permittivity and magnetic permeability respectively. Here, ${\omega _p}$, ${\omega _{\rm{ce}}}$ and ${v_{c}}$ are plasma frequency, electron cyclotron frequency and electron collision frequency, respectively.
The background magnetic field is $\mathbf{B_{0}} = B_{y}\hat{y} + B_{z}\hat{z}$. The wave propagates with magnetic inclination angle ${\theta}$ with respect to the ${\mathbf{z}}$ axis. Equation (3) is only for cold plasmas and the ion motion is neglected.

%Figure 1 shows the EM wave propagation model in plasmas with arbitrary magnetic inclination angle ${\theta}$  with respect to the ${\mathbf{z}}$ axis. The electromagnetic wave is propagating along the ${\mathbf{z}}$ axis direction. The scattered field for a 1-D magnetized plasma slab are solved based on the current density convolution finite-difference time domain (JEC-FDTD) \cite{b11}.

To verify our method, we calculate the scattering field $\mathbf{E_{\rm s}}$ for a one-dimensional magnetized plasma slab shown in Fig.3. The computational space takes up 800 grid along the ${\mathbf{z}}$ directions, among which plasma takes up 200-600 grids with uniform distribution and a thickness of $d =3 \, \rm{cm}$. The JEC-FDTD method is applied to obtain the solution of Equation(9). The algorithm can be found in details \cite{b11}. For uniform plasma distribution, parameters in the simulation are set as following in Table I.

\begin{table}[h]
\caption{System Parameter}
\label{table}
\center
\setlength{\tabcolsep}{2pt}
\begin{tabular}{|p{150pt}|p{50pt}|}
\hline
\rule{0pt}{10pt}
Symbol &
Value \\
\hline
\rule{0pt}{10pt}
%Space iteration step size&
%$ dz: 75 $ um  \\
%\hline
%\rule{0pt}{10pt}
%Time iteration step size&
%$ dt: 0.125 $ ps  \\
%\hline
%\rule{0pt}{10pt}
Electron collision frequency $\nu_{c}$&
$  20 $ GHz  \\
\hline
\rule{0pt}{10pt}
Electron cyclotron frequency $\omega_{ce}$&
$  30 $ GHz  \\
\hline
\rule{0pt}{10pt}
Uniform plasma frequency $\omega_{p}$&
$  40 $ GHz  \\
\hline
\rule{0pt}{10pt}
Magnetic inclination angle $\theta$ &
$ 90\,\degree $ \\
\hline
\end{tabular}
\label{tab1}
\end{table}

%333333333333333333333333333333333
%333333333333333333333333333333333
%333333333333333333333333333333333
\begin{figure}[htbp]
\centering
\includegraphics[width=0.4\textwidth]{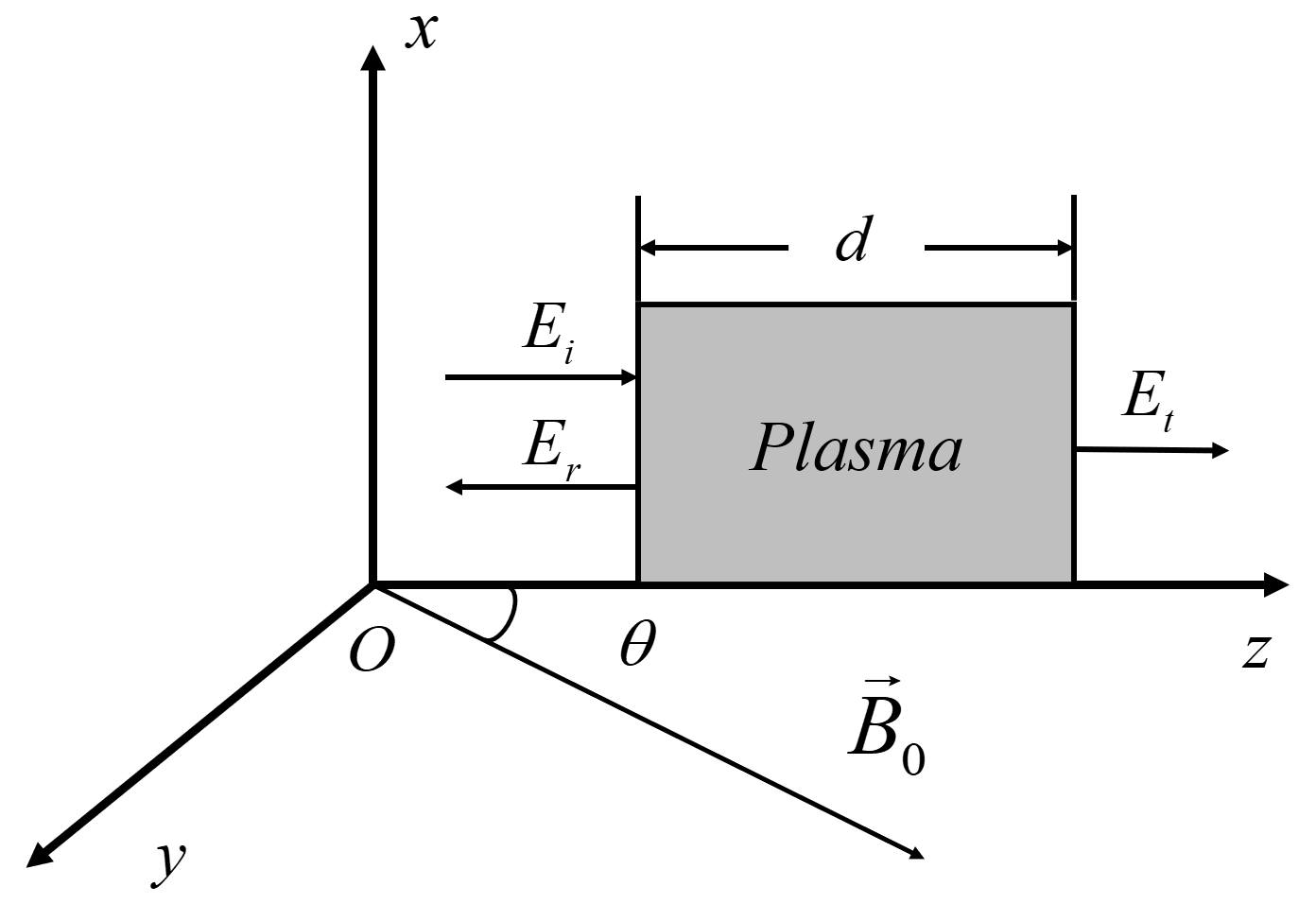}
\caption{Multiple-physics system: EM wave interaction with plasmas  with arbitrary magnetic inclination $\theta$. The spatial domain is in the ${\mathbf{z}}$ direction.}
\end{figure}

For $\theta= \pi/2$, Equations (10) are expressed as a scalar differential equation as:
\begin{equation}
\dfrac{\partial E_{x}}{\partial z}=-\mu_{0}\dfrac{\partial H_{y}}{\partial t}
\end{equation}
\begin{equation}
-\dfrac{\partial H_{y}}{\partial z}=\varepsilon_{0}\dfrac{\partial E_{x}}{\partial t}+J_{x}
\end{equation}
\begin{equation}
\dfrac{\partial J_{x}}{\partial t}+\nu_{c}J_{x}=\varepsilon_{0}\omega_{p}^{2}E_{x}+\omega_{ce}J_{z}
\end{equation}
\begin{equation}
\dfrac{\partial J_{z}}{\partial t}+\nu_{c}J_{z}=\varepsilon_{0}\omega_{p}^{2}E_{z}-\omega_{ce}J_{x}
\end{equation}
\begin{equation}
J_{z}=-\varepsilon_{0}\dfrac{\partial E_{z}}{\partial t}
\end{equation}

Here, there are 5 Equations to discover with five unknown parameters $\mu_{0}, \varepsilon_{0}, \varepsilon_{0}\omega_{p}^{2}, \omega_{ce}, \nu_{c}$. This set of PDE equations is our first goal for equation discovery. This set of PDE equations is for EM wave propagation in anisotropic plasmas, however, it can effective reveal the application for our proposed network.

\subsection{Prepossessing}
The whole process of inverse model is shown in Fig. 4. First of all, the physical input data are $E_{x}, J_{x}, E_{z}, J_{z}, H_{y}$ calculated from JEC-FDTD algorithm. Then, sampling is conducted for the input data $ E_{x}, J_{x}, E_{z}, J_{z}, H_{y}$ for training, respectively. 

Next, normalization is executed directly after data sampling. As we use the same input for different equations, we need to do normalization for each equations because the order of magnitudes of input data will cause big error for data training and lead to wrong results. Specifically, we multiply each input data with a designed parameter respectively to keep them into the same order of magnitude. Afterwards, each input data should be compared with time derivative $\frac{\partial E_{x}}{\partial t},\frac{\partial J_{x}}{\partial t},\frac{\partial E_{z}}{\partial t},\frac{\partial J_{z}}{\partial t}, \frac{\partial H_{y}}{\partial t}$, which are included by the target equation to keep in the same magnitude.

%4444444444444444444444444444444444444
%4444444444444444444444444444444444444
%4444444444444444444444444444444444444

\begin{figure}[htbp]
\centering
\includegraphics[width=0.3\textwidth]{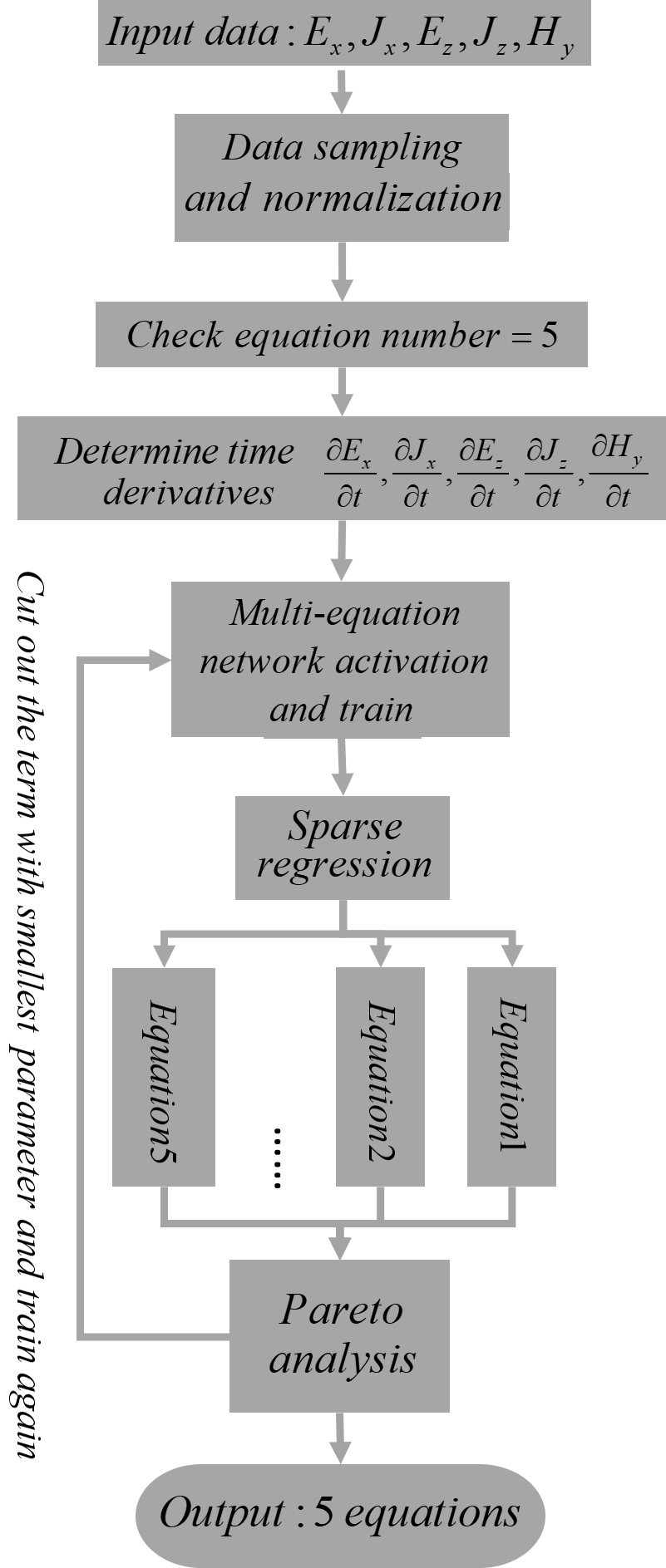} 
\caption{The overall process of the inverse model for a multiple-physics EM problem. Step 1: Data collected from a solution of PDE Equations (11-15). Step 2: Data sparse sampling and normalization. Step 3: Check the  number of equations and determine the time derivative of the input data. Step 4: Multi-Equation network activation and training shown in Fig.1. Step 5: Sparse regression is set at the end of training and the detailed form of equations are shown. Step 6: After Pareto analysis, distortion terms will be deleted.}
\end{figure}

For collecting training data  we select three physical quantity of the scattered field $ E_{x}, E_{z}, H_{y}$ and two currents $J_{x}, J_{z}$ at spatial grids $z(\Delta z)=190\sim210$ including plasma-vacuum boundary (boundary grid $200$). For temporal series of physical quantities, we choose few samples at the time period $t(\Delta t)=500\sim510$, which includes the maximum value of the scattered field from plasmas. 

%55555555555555555555555555555555555
%55555555555555555555555555555555555
%55555555555555555555555555555555555
\begin{figure}[htbp]
\centering
\includegraphics[width=0.50\textwidth]{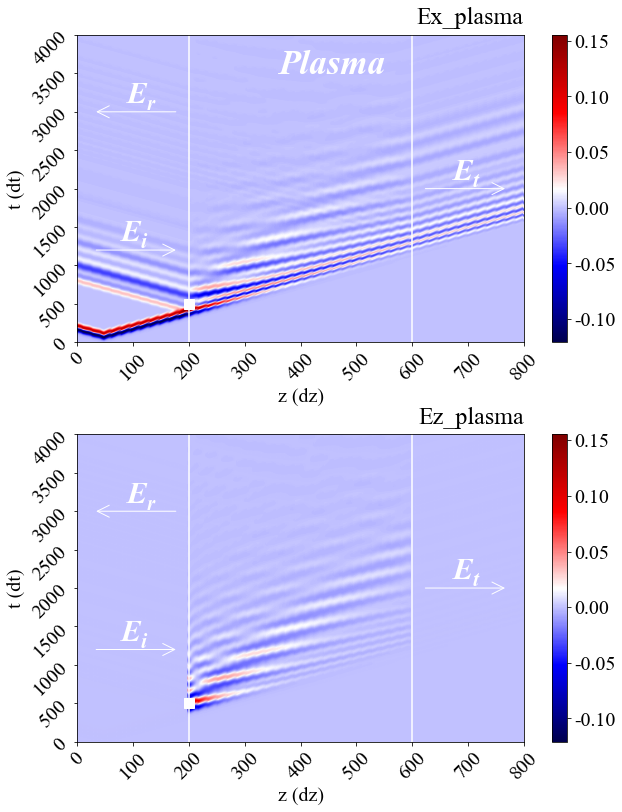}
\caption{Data collection location - for upper part, this is a x-direction electric field intensity $ E_{x} $ picture. The two white lines means the incident location and exit location. The small white square represents the collection area. For lower part, it is the all points collected, where the grey line represents the incident border of plasma. We captured points both inside and outside plasma to ensure the reliability.}
\end{figure}

Figure 5 shows examples for the total electric field $E_{x}$ and $ E_{z}$ in perpendicular to the magnetic field. The sampling scheme is the same for all five quantities. In Figure 5, the two white lines represent the boundary of plasma. The small white square represents the data collection area from both inside and outside the plasma. For time sampling, we choose 10 time steps after time 500, because the gradient of curve at this time is obvious so that the function is easy to be recognized by neural network.

%66666666666666666666666666666666666
%66666666666666666666666666666666666
%66666666666666666666666666666666666
\begin{figure}[htbp]
\centering
\includegraphics[width=0.50\textwidth]{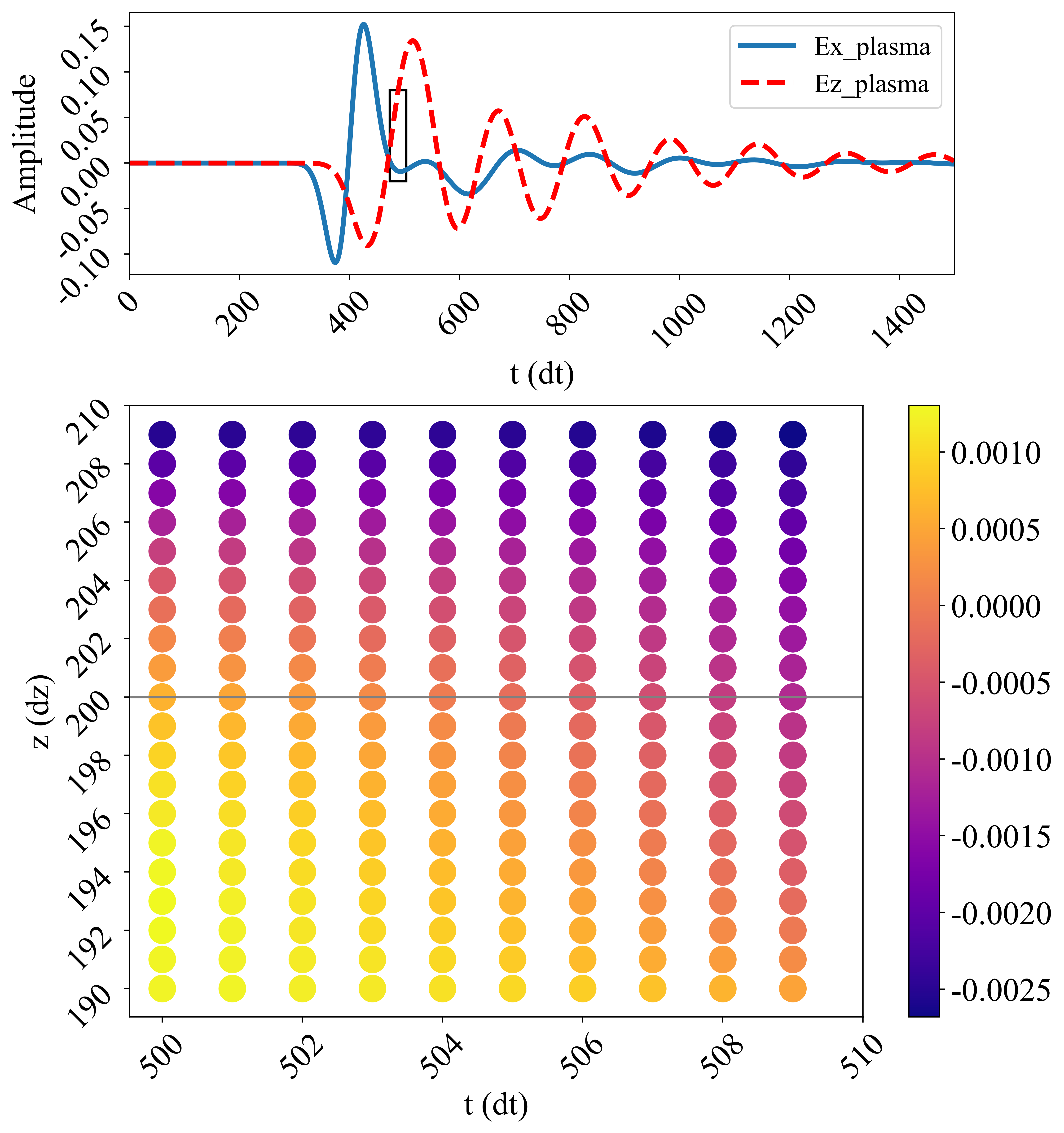}
\caption{Sampling area - in the upper panel, the black square shows the data sampling area for $E_{x}$ and $E_{z}$ in time domain. In the bottom panel, the black horizontal line indicates the border of plasma and our collection area begins right on the incident border.}
\end{figure}

%for upper part, this is a x-direction electric field intensity $ E_{x} $ picture. The two white lines means the incident location and exit location. The small white square represents the collection area. For lower part, it is the all points collected, where the grey line represents the incident border of plasma. We captured points both inside and outside plasma to ensure the reliability.

The sampling area for training network is shown in Figure 6. In the upper panel, the black square shows the data sampling area for $E_{x}$ and $E_{z}$ in time domain. In the bottom panel, the black horizontal line indicates the border of plasma and our collection area begins right on the incident border. It shows that only a small number of time and spatial information are continuously sampled for training.

\subsection{Network Construction}
From forward Euler's method, we have
\begin{equation}
\tilde{E_{x}}(t_{i+1},\cdot)=E_{x}(t_{i},\cdot)+\Delta t\cdot\dfrac{\partial E_{x}}{\partial t}
\end{equation}
\begin{equation}
\tilde{H_{y}}(t_{i+1},\cdot)=H_{y}(t_{i},\cdot)+\Delta t\cdot\dfrac{\partial H_{y}}{\partial t}
\end{equation}
\begin{equation}
\tilde{J_{x}}(t_{i+1},\cdot)=J_{x}(t_{i},\cdot)+\Delta t\cdot\dfrac{\partial J_{x}}{\partial t}
\end{equation}
\begin{equation}
\tilde{J_{z}}(t_{i+1},\cdot)=J_{z}(t_{i},\cdot)+\Delta t\cdot\dfrac{\partial J_{z}}{\partial t}
\end{equation}
\begin{equation}
\tilde{E_{z}}(t_{i+1},\cdot)=E_{z}(t_{i},\cdot)+\Delta t\cdot\dfrac{\partial E_{z}}{\partial t}
\end{equation}

For one-dimension problem, the spatial direction is only $ z $, so we use only one filter to get one kind of spatial differentiation. Take $ E_{x} $ for example:
\begin{equation}
h_{10}\otimes E_{x}\approx \dfrac{1}{2}\delta_{z}\dfrac{\partial E_{x}}{\partial z}
\end{equation}

%66666666666666666666666666666666666666
%66666666666666666666666666666666666666
%66666666666666666666666666666666666666
%\begin{figure}[htbp]
%\centering
%\includegraphics[width=0.5\textwidth]{04-2.png}
%\caption{The adjusted neural network structure - function $ h $ %contains all possibilities of this equation and as the spatial %direction is only z, so we use only one filter to get one kind of %spatial differentiation.}
%\end{figure}

Based on this, we get the adjusted neural network structure. To fulfil the function of Equation (21), we use a funtion called conv2d in TensorFlow to generate convolution calculation. In our case , since it is one dimensional 1D problem, we obtain the shape of convolution kernel of $ 2\times1 $, so that this function still works.

By screening different weights of network, the correct equation term and its corresponding coefficient can be obtained. This is one single time block for Equations (11-15).
  
As one single block $\delta_{t}$ given above, we have 10 $\delta_{t}$ time block to build up the whole network. By using the parameter sharing characteristic of neural network, the training coefficient results become accurate if the number of time blocks is increased. Consider that all redundant terms may lead to the problem of complicated neural network structure, we reduce the number of candidates based on certain prior physical knowledge, thus making the network training more efficient.

%It is clear that only a small number of information and the distance of each spatial point is very close. This character is quite meaningful in radar communication when ground detection location is limited.

\begin{table*}[htbp]
%\makeatletter\def\@captype{table}\makeatother
\caption{Inversion Result for PDEs with 10 spatial points for each equation}
\label{table}
\centering
\setlength{\tabcolsep}{5pt}
\begin{tabular}{|p{130pt}|p{170pt}|p{170pt}|}
\hline
\rule{0pt}{15pt}
Equation& 
Error without noise&
Error with noise SNR = 40$\rm{dB}$ \\
\hline
\rule{0pt}{18pt}
$ \dfrac{\partial E_{x}}{\partial z}=-\mu_{0}\dfrac{\partial H_{y}}{\partial t} $ &
$ \mu_{0}: 1.68\% $ 

&
$ \mu_{0}: 10.39\% $ \\
\hline
\rule{0pt}{18pt}
$ -\dfrac{\partial H_{y}}{\partial z}=\varepsilon_{0}\dfrac{\partial E_{x}}{\partial t}+J_{x} $ &
$ \varepsilon_{0}:10.14\% $ (the average of two terms)

&
$ \varepsilon_{0}:12.92\% $ (the average of two terms) \\
\hline
\rule{0pt}{18pt}
$ \dfrac{\partial J_{x}}{\partial t}+\nu_{c}J_{x}=\varepsilon_{0}\omega_{p}^{2}E_{x}+\omega_{ce}J_{z} $ &
$ \nu_{c}: 9.12\%,\: \varepsilon_{0}\omega_{p}^{2}: 0.53\%,\: \omega_{ce}: 1.06\% $ 
&
$ \nu_{c}: 23.38\%,\: \varepsilon_{0}\omega_{p}^{2}: 6.84\%,\: \omega_{ce}: 3.07\% $\\
\hline
\rule{0pt}{18pt}
$\dfrac{\partial J_{z}}{\partial t}+\nu_{c}J_{z}=\varepsilon_{0}\omega_{p}^{2}E_{z}-\omega_{ce}J_{x} $ &
$ \nu_{c}: 0.39\%,\: \varepsilon_{0}\omega_{p}^{2}: 1.02\%,\: \omega_{ce}: 1.46\% $ 
&
$ \nu_{c}: 9.19\%,\: \varepsilon_{0}\omega_{p}^{2}: 5.73\%,\: \omega_{ce}: 35.14\% $ 
\\
\hline
\rule{0pt}{18pt}
$ J_{z}=-\varepsilon_{0}\dfrac{\partial E_{z}}{\partial t} $ &
$ \varepsilon_{0}:4.32\% $  &
$ \varepsilon_{0}:5.33\% $
\\
\hline
\end{tabular}
\label{tab1}
\end{table*}

\subsection{Result and Analysis}
%More space points can make the prediction results more accurate, but in practical application, that is, in the process of radar communication, it is difficult for us to collect data in a wide range of space, most of which come from a few points in the center. Therefore, we train this network for different cases of different spatial points number.
For the system of PDEs set, we build a neural network based on sparse temporal-spatial sample data for training and predict the coefficients of PDEs without interfering each other. In this process, we need to fully normalize the data, taking into account that the magnitude of various field quantities of electromagnetic field data with huge differences. By evaluating the order of magnitude of each candidate, we set a reasonable normalization coefficient for each candidate with certain prior knowledge to ensure that the disparity of magnitude order will not cause great impact on training results of the network. 

\subsubsection{Homogeneous simulation}
Wave propagation in homogeneous medium is the most common problem for EM applications. Therefore, we select a homogeneous case as our first attempt to validate methodology proposed above.

Table II summarizes inversion results of proposed methodology applied to a multiple-physics EM interaction system of Equations (11-15) with all coefficients homogeneous. All data shown in Table II are sampled from $ z=200 $ to $ z=209 $, and $ t=500 $ to $ t=509 $. In total, 10 spatial points and 10 time points are selected for comparison. The five PDEs are obtained simultaneously with five coefficients $\mu_{0}, \varepsilon_{0},\nu_{c}, \varepsilon_{0}\omega_{p}^{2}, \omega_{ce}$. It is noted that $\nu_{c},\omega_{p}, \omega_{ce}$ are three important  parameters of collisional magnetized plasmas.  For simplicity, we combine $ \epsilon_{0}\omega^{2}_{p} $ together as one parameter. Most of the relative errors are relatively low approximately less than $5\%$. We add noise into the input data with signal to noise ratio $\rm{SNR}=40 \,\rm{dB}$. The equations with some noise $\rm{SNR}=40 \,\rm{dB}$ can also be identified with increased errors. In general, the inversion error increases but can still discover physical system even if significantly sub-sampled spatially and temporally. 
%77777777777777777777777
%77777777777777777777777
%77777777777777777777777
\begin{figure}[htbp]
\centering
\includegraphics[width=0.45\textwidth]{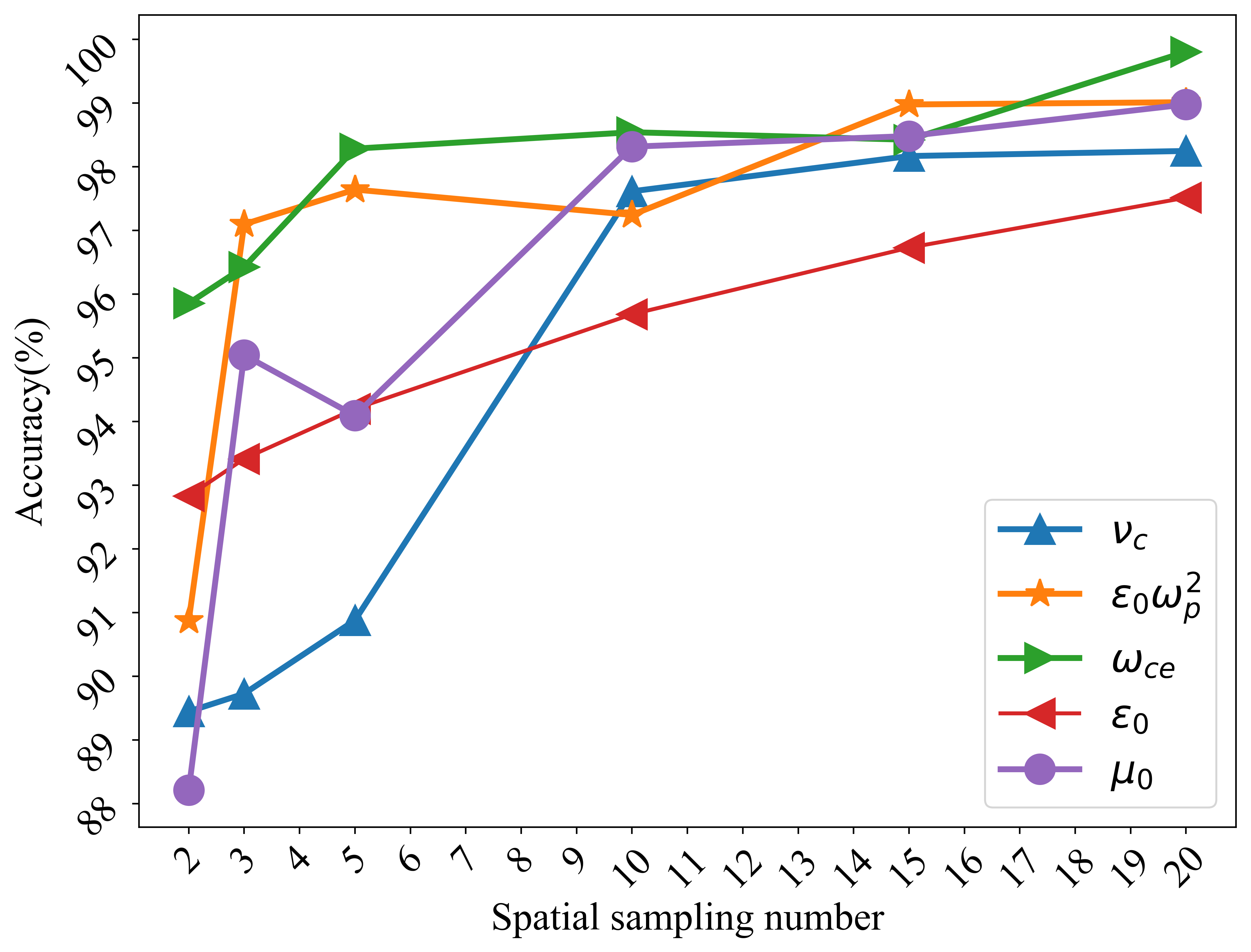} 
\caption{Inversion accuracy dependent on spatial sampling numbers. It is noted that each point indicates the accuracy of normalized parameters. The spatial sampling points are 2, 3, 5, 10, 15, 20, respectively and the beginning location is $z=200 $. It is noted that inversion accuracy reaches to $95\%$ only for three samples but reduces substantially for two samples.}
\end{figure}

The sampling scheme of physical quantities in spatial and time domain is the most important parameter for calculating derivatives to ensure inversion accuracy. On one hand, the dependence of inversion accuracy on spatial sampling numbers is depicted for coefficients $\nu_{c}, \varepsilon_{0}\omega_{p}^{2}, \omega_{ce}, \varepsilon_{0}, \mu_{0}$ in PDEs in Figure 7. The spatial sampling points are 2, 3, 5, 10, 15, 20, respectively and the beginning location is $z=200 $ at  plasma-vacuum boundary. It is noted that inversion accuracy reaches to $95\%$ for three spatial sampling position but reduces substantially for just two spatial samples. Thus, for considering the spatial sampling limitation, we need to make trade-off between accuracy and spatial sampling scale.  

On the other hand, temporal sampling scenario of the field sequence data plays an important role in inversion accuracy as well. Figure 8 depicts the dependence of accuracy $\varepsilon_{0}$ on the gradient of $E_{\rm z}$ based on Equation (15) as an example. We select three typical time sampling blocks corresponding to low (red), middle (blue), and high gradient (green) in the upper panel of Figure 8. In each block, there are 10 temporal samples. It is noted that the accuracy reaches $99\%$ for high gradient (green) and reduces to $90\%$ for low gradient (red). It is noted that the inversion accuracy correlates with the time derivative of the sampled data.  With data varying rapidly with time, the neutral network can catch  data characteristics easily. For complete PDE equation set, it will require a common time block for all quantities with high gradients within reasonably large amplitude.       

%888888888888888888888888
%888888888888888888888888
%888888888888888888888888
\begin{figure}[htbp]

\includegraphics[width=0.5\textwidth]{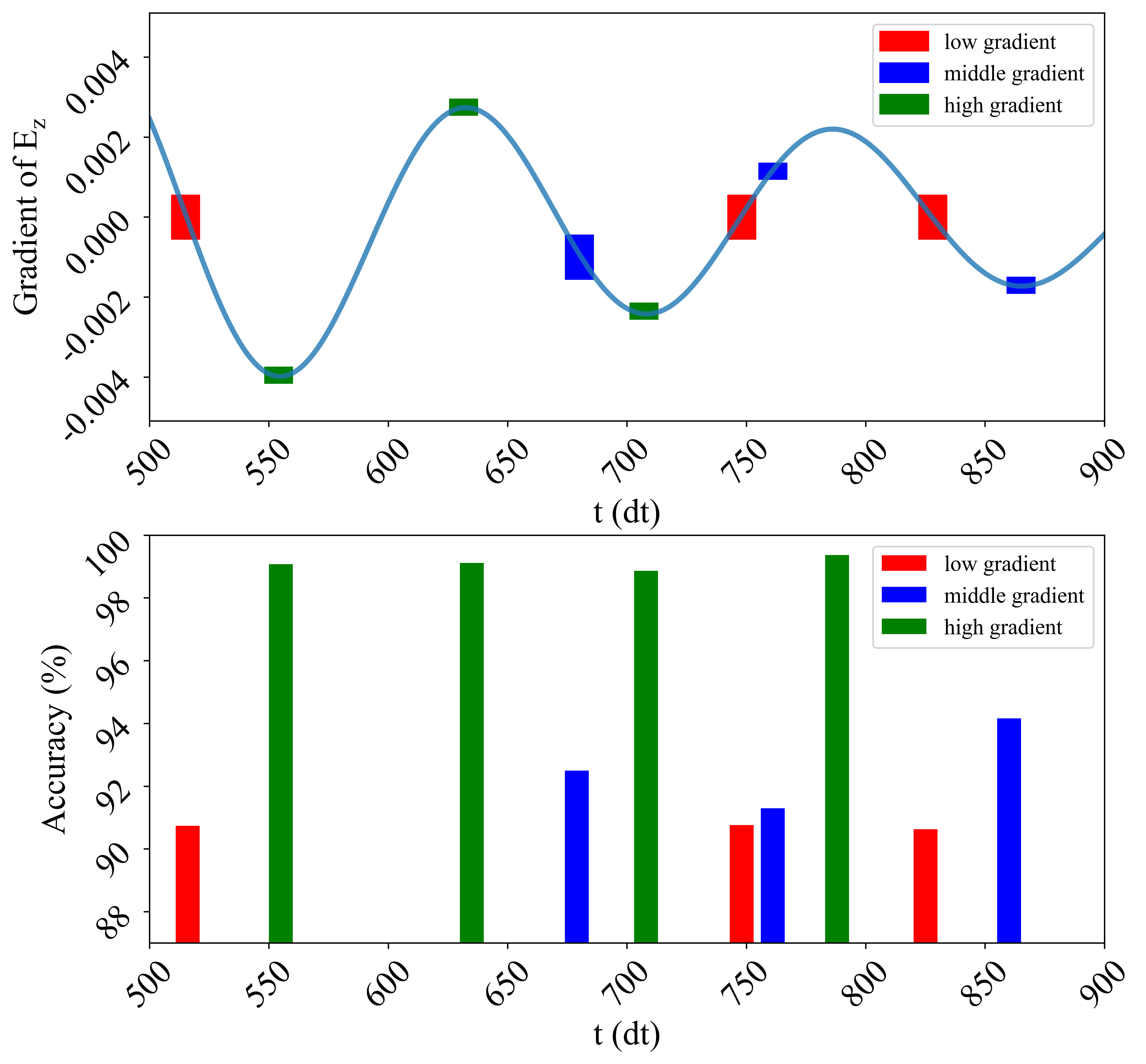} 
\caption{Inversion accuracy dependent on temporal sampling scenario. Three time sampling blocks are marked for low (red), middle (blue), and high gradient (green) with 10 samples in each block. Here is an example of Equation (15) with the time derivative of $ E_{z} $ at spatial location $ z=200 $. For low gradient (red block in upper panel), the accuracy is low (red in lower panel). It is noted that the inversion accuracy correlates with the time derivative of sampled data block.  }
\end{figure}

\subsubsection{Inhomogeneous simulation}
For all simulation above, we assume homogeneous coefficient for discovering the PDEs set. To verify algorithm further, we investigate the inhomogeneous coefficient further which often occurs for a wide range of applications. 

%神经网络调整 - 神经网络的构架需要被调整以适应非均匀的问题. 由于非均匀项的系数是随着空间位置一直在发生变化，因此我们需要把方程的一部分系数（也就是存在非均匀系数项的系数）从一个单一的数字调整成一个长度可调整的张量。我们把这些长度可调整的张量称作“非均匀系数张量”，这个张量将会在神经网络中被训练。非均匀系数张量的长度是由空间采样点数的多少决定的，一般与空间采样点数保持一致，同时，我们假设哪些项的系数非均匀的这一点对我们来说是已知的，那么这就可以不必让所有的候选项的系数都变成非均匀系数张量，从而使得训练过程得以简化。这个非均匀系数张量也就是表征了该非均匀系数在这个狭小的采样空间内系数的变化情况。在完成训练后，我们将得到的非均匀系数张量取出，并利用梯度下降法将它拟合为一个简要的函数表达式形式，也就是这个非均匀系数的变化规律了。

Network adjustment. The structure of neural network needs to be adjusted to solve inhomogeneous problems. First, as the coefficients of inhomogeneous terms are varying with spatial position, we change parameters of inhomogeneous terms from a single number to a length-adaptive tensor. We call this length-adaptive tensor \textit{inhomogeneous parameter tensor}. This tensor will be trainable in neural network. The length of inhomogeneous parameter tensor is equal to the spatial sampling point number. Secondly, we assume that prior knowledge is known which candidate terms have inhomogeneous parameters so that not all candidate terms should have inhomogeneous parameter tensors. 
Finally, the predicted parameter is no longer a single number but a tensor which indicates the varying parameters within spatial domain. After training, the underlying equations with inhomogeneous coefficients can be distilled from the inhomogeneous parameter tensor by the gradient descent method. 

The input data of the scattered field for inhomogeneous density  is obtained based on the JEC-FDTD algorithm. Here, we assume that $\omega_{p}^{2}$ obeys a sinusoidal envelope, positively proportional to plasma density for inversion. A normal expression of varying parameter can be written as
\begin{equation}
\omega_{p}^{2}\sim A\sin(kz+\varphi)+C
\end{equation}
Here, $A, k, \varphi, C$ is amplitude, wave number, phase and constant, respectively. Plasma parameter $\omega_{p}^{2}$ varies as Equation (22) from $z=200$ to $z=600$ with two wavelength inside plasmas correspondingly. 

In our simulation, the exact form and normal expression of inhomogeneous parameter are unknown. In general, we assume that unknown inhomogeneous coefficients can be expressed as a kind of trigonometric series in the form of
\begin{equation}
\dfrac{1}{2}A_{0}+\sum_{n=1}^{N}(A_{n}\sin (nx)+B_{n}\cos (nx))
\end{equation}
where $n$ is the order of the series, $A_{n}$ and $B_{n}$ are coefficients, respectively.

We assume that $ N = 3 $ is sufficient to represent the expression of inhomogeneous parameter tensor. Thus, the estimated function of inhomogeneous parameters  $\tilde{h}(z)$ can be expressed as  
\begin{equation}
\begin{split}
\tilde{h}(z)=A_{1}\sin (kz) + B_{2}\cos (kz) + A_{2}\sin (2kz) +\\
B_{2}\cos (2kz) + A_{3}\sin (3kz) + B_{3}\cos (3kz) + C
\end{split}
\end{equation}
where $A_{1}, A_{2}, A_{3}$ and $B_{1}, B_{2}, B_{3}$ are unknown coefficients to discover, respectively.  

All unknown coefficients are determined in Equation (24) by  sparse regression. We use adaptive moment optimizer (Adam optimizer) to solve the following equation 
\begin{equation}
\begin{split}
\arg\min_{\mathbf{S}}\sum\Vert h(z)-\tilde{h}(z)\Vert_{2}+\lambda\Vert{\mathbf{S}}\Vert_{1} \\
{\mathbf{S}}=[A_{1}, B_{1}, A_{2}, B_{2}, A_{3}, B_{3}, C]
\end{split}
\end{equation}

For spatial sampling, we choose 20 spatial position collected from $z=240-260$ for training and then predict inhomogeneous coefficient in the whole spatial domain. The identified coefficients of ${\mathbf{S}}$ are $A_{1}=0.5036, B_{1}=-0.0011$, $A_{2}=-0.0020, B_{2}=0.0002$ and $A_{3}=0.0005, B_{3}=0.0018$, respectively. The identified constant number is $C=0.9951$ with true value 1. The dominant term $A_{1}\sin(kz)$ is successfully selected among all candidates and other terms are small enough to be ignored. The true and identified models of inhomogeneous coefficient are given by 
\begin{equation}
\omega_{p}^{2}\sim 0.5\sin(3.142z)+1 \,\rm{(True)}
\end{equation}
\begin{equation}
\omega_{p}^{2}\sim 0.5036\sin(3.123z)+0.9951\, \rm{( Identified)}
\end{equation}

Figure 9 shows the inversion result of inhomogeneous coefficient in comparison with theoretical value by Equation (26). The spatial sampling position is shown by the black box from $z=240-260$ for training. Although sampling only at few spatial grids, inhomogeneous coefficients can be obtained in the sampling regime and then make prediction outside measurement regime according to discovered equations . Basically, the inversion result agrees well with theoretical value after iteration with only 20 spatial and 10 temporal samples. For a Gaussian white noise
of $\rm{SNR}=40\rm{dB} $, the identified model with noise is
\begin{equation}
\omega_{p}^{2}\sim 0.5897\sin(3.360z)+1.082\, \rm{( Identified)}
\end{equation}

However, there are some errors of prediction with noise. One possible source of error comes from the neural network training of inhomogeneous parameter tensor. Another source of error may arise from processes during distilling the underlying equation from inhomogeneous parameter tensor. Even though, this method still discover and predict the inhomogeneous parameter.   

%9999999999999999999999999

%9999999999999999999999999
\begin{figure}[htbp]
\centering
\includegraphics[width=0.5\textwidth]{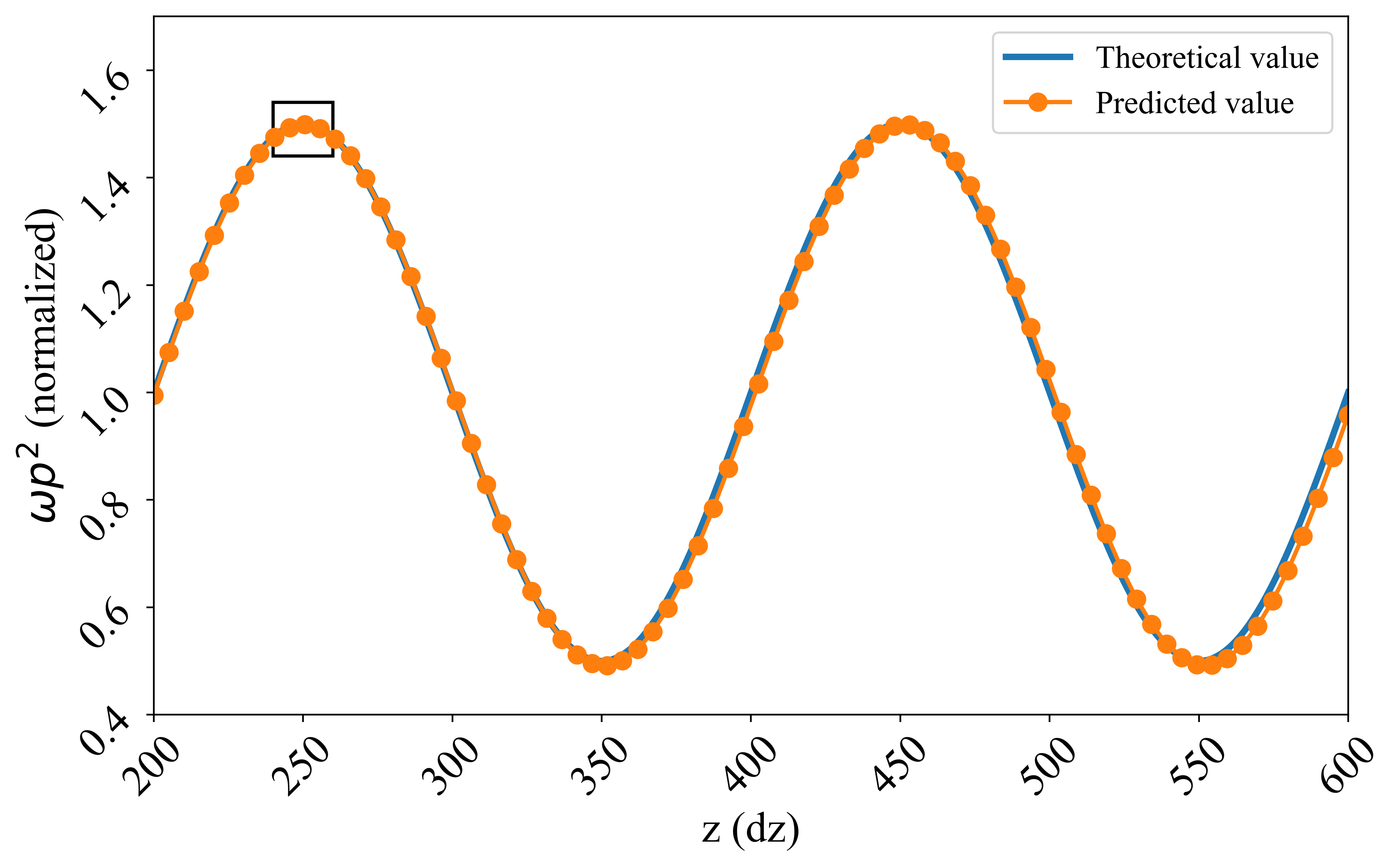} 
\caption{Inhomogeneous parameter prediction without noise. The blue solid line is theoretical value, and the dotted line is predicted value. The black box indicates spatial sampling position from $z=240-260$ inside plasmas for training.}
\end{figure}

%We can see from the figure() that even only use 5 point which means very close to an isolated monitoring site that is such results have some value for practical application. We add Gaussian white noise in signals for the system anti noise test, here it is important to note that due to the time difference for anti-noise performance is very poor, the main reason is that its change is very severe, slight amounts of noise may produce a great influence on the essence of the signal, so that the prediction results serious distortion.
%
%Often our training results will have many redundant terms, that is, other candidates also have a weak coefficient. By using Pareto analysis, we can remove these distortion terms effectively.

\section{Conclusion and Discussion}
In this paper, we have presented a a data-driven network architecture to discover the hidden nonlinear PDE equation set, where first-principles derivation may be intractable. The architecture extends the idea from the PDE-net with compressive sensing, which approximates differential operations by convolutions with properly constrained filters and approximate the nonlinear response by deep neural networks. We add a coefficient regression module to the system and Pareto analysis to improve the identification accuracy. The architecture can contribute construct a model which characterizes the observed dynamics and generalizes to unsampled regime of parameter space.
 
We apply this network architecture for Maxwell’s equations for a multiple-physics EM system. The discovery of PDEs are complex in terms of multiple-physics, non-linearity, inhomogeneity, and anisotropy. We can discover Maxwell's equations coupled with plasma current conservation equation with low error on the coefficient of PDEs. The homogeneous coefficients of PDEs have been constructed based on sparse temporal-spatial data. Also, our attempt demonstrates the possibility of inferring inhomogeneous coefficients of physical parameters with prior knowledge. By assuming that the expression of coefficient of inhomogeneous parameters can be equal to some order of trigonometric series, we use sparse regression to discover the equation underlying in the inhomogeneous coefficients with little prior knowledge. The identified equation requires a batch of adjacent temporal-spatial samples to evaluate derivatives of physical variables. When the discretized solution contains measurement noise, the numerical derivatives becomes challenging to evaluate that will need further work. 

Finally, it is noted that current network has been tested for the 1-D PDEs  for the extraordinary wave $\theta=90^{\rm{o}}$ with only 5 equations coupled. Considering arbitrary angle $\theta$ in 3-D spatial domain, we need to add the representation module of curl operator in the neural network. Secondly, this method has been simplified to identify models with some prior knowledge for inhomogeneous parameter. If no prior knowledge is available, we will need a new candidate function database for identifying unknown terms similary as homogenous network. Thirdly, kinetic effects of plasmas are ingnored in the forward model, and we make assumputions for cold plasmas as well. Finnaly, it is valuable to try the proposed framework on real data of nonlinear EM wave and space plasma interaction experiments\cite{b12,b13,b14} and other potential applications.

\section*{Acknowledgement}
This work was supported by the National Key Research and Development Program of China under Grant 2017YFB0502703 and also supported by the Shanghai Science Foundation under Grant 19ZR1403900 and Grant 19511100600.

\end{document}